\documentclass{aa}

\usepackage{graphicx}
\usepackage{txfonts}
\usepackage{epsfig,graphics,graphicx,bm,amssymb}
\begin{document}

\title{A new scenario for the origin of the 3/2 resonant  system HD\,45364}

\author{J. A. Correa-Otto
          \inst{1}\fnmsep\thanks{e-mail: jorge9895@usp.br} \and
          T. A. Michtchenko\inst{1}
          \and C. Beaug\'e\inst{2} }

\institute{Instituto de Astronomia, Geof\'isica e Ci\^encias Atmosf\'ericas, USP, Rua do Mat\~ao 1226, 05508-090 S\~ao
Paulo, Brazil\\
         \and
             Instituto de Astronom\'ia Te\'orica y Experimental, Observatorio Astron\'omico, Universidad Nacional de
C\'ordoba, Laprida 854, (X5000BGR)
C\'ordoba, Argentina\\
             }

   \date{Submitted to A\&A. Revised version: 20/09/2013}


\abstract{ In this paper we revise the model proposed by Rein et al. (2010) for the origin of the HD\,45364
exoplanetary system, currently known to host two planets close to the 3/2 mean-motion commensurability (MMR). We show
that, due to high surface density of the protoplanetary disk needed for Type III migration, this model could only lead
to planets in a quasi-resonant regime of motion, and thus not consistent with the resonant configuration obtained by
Correia et al. (2009). Although both resonant and quasi-resonant solutions are statistically indistinguishable with respect to radial velocity measurements, their distinct dynamical behaviour is intriguing. We use the semi-analytical model to confirm the quantitative difference between two configurations. In order to form a
system evolving \textit{inside} the 3/2 resonance, we develop a different model. Our scenario includes an interaction
between different (but slower) planetary migration types, planet growth, and gap formation in the protoplanetary disk.
The choice of the described evolutionary path was due to a detailed analysis of the structure of the phase space in the vicinity of the 3/2 MMR employing dynamical mapping techniques. The outcomes of our simulations are able to reproduce very closely the 3/2 resonant dynamics obtained from the best-fit presented by Correia et al. (2009). In addition, varying the strength of the eccentricity damping, we can also simulate the quasi-resonant configuration similar to that in Rein et al. (2010).  We furthermore show that our scenario is robust with respect to the physical parameters involved in the resonance trapping process. However, the confirmation of our scenario will be possible only with additional radial velocities measurements.  }


   \maketitle
%

\section{Introduction.}

HD\,45364 is the first discovered candidate with two planets evolving inside or near the 3/2 mean-motion resonance (MMR)
(Correia et al. 2009). Recently, the HD\,204313 system was also suggested as a candidate for this resonance (Robertson
et al. 2012), but the stability of the proposed configuration still needs to be confirmed.  There are also
several Kepler systems located near the 3/2 resonance, however, their involvement into the resonance is uncertain due
to the lack of information on their eccentricities. The long-term survivance of the  HD\,45364 system in the 3/2 MMR
corresponding to the best-fit of Correia et al. (2009) was tested and confirmed through numerical integrations of the
exact equations of motion (Correia et al. 2009).

The semimajor axes of the HD\,45364 planets \textbf{b} and \textbf{c} are equal to 0.682\,AU and 0.898\,AU,
respectively. A study of the origin of such compact planetary configuration was first carried out by Rein et al.
(2010). To form this resonant pair, the authors initially tested  the standard scenario of a convergent migration of
two planets embedded in a protoplanetary disk. They noted, however, that a Type II migration most likely leads to a
capture inside the strong 2/1 MMR, which formed a dynamical barrier that blocked the road to the 3/2 MMR. To overcome
this problem, the authors suggested an alternative scenario, introducing a non-conventional Type III migration.  Due to the high density of the disk assumed, this type of migration produces a very rapid orbital decay of the planets, allowing them to by-pass the 2/1 resonance and capture in the 3/2 MMR.

However, one fact, already noted by Rein et al. (2010), attracted our attention: a discrepancy in the dynamical
behaviour between the simulated solution and that obtained from the best-fit of Correia et al. (2009). The difference
lies mainly in the magnitudes of the planet eccentricities (it was explained by the authors as due to possible errors
in the radial velocity data). Moreover, other discrepancies appear when the dynamics of the Rein et al. (2010) solution is analyzed in detail. A N-body simulation shows that the proposed solution does not lie within the libration zone of
the 3/2 MMR, but exhibits a quasi-resonant behavior. In addition, some characteristic dynamical quantities, such as
frequencies and amplitudes of oscillation of orbital elements, differ by orders in magnitude when compared to those of
the best-fit solution.

It should be noted that, notwithstanding the distinct dynamical behaviour of two solutions, both solutions lead to practically the same radial velocity measurements; thus from an observational point of view, the two solutions are
statistically indistinguishable. Only additional radial velocity measurements will allow us to unambiguously determine
the resonant state of the system. HD45364 is therefore an ideal test case for planet formation scenarios.

To form a pair of planets evolving \textit{inside} the 3/2 resonance, we consider a different scenario in this paper. Although we also suggest resonance trapping following planetary migration from disk-planet interactions, there are some important differences: (i) we assume a slower migration rate, as predicted by either Type I or Type II migration modes; (ii) we adopt different disk parameters, especially, the lower surface density; (iii) we introduce mass growth during both the migration and resonance trapping processes, and (iv) we assume that the migration process begins when the planets are in the embryonic stage ($\sim 0.1$\,M$_\oplus$) and thus much smaller than the present day Saturn-size planets.

We show that, depending on just one parameter, our scenario is able to reproduce the averaged magnitudes of
the orbital elements and the resonant behaviour of the best-fits of Correia et al. (2009), also as the solution given in Rein et al. (2010). In addition, we perform the study of the phase space of the 3/2 mean-motion resonance
using a semi-analytical model originally developed by Michtchenko et al. (2008a). This analysis allows us to compare the dynamical behavior of the best-fit solution of Correia et al. (2010) and the solution obtained by Rein et al. (2010), and identify the most significant differences between them.

This paper is organized as follows. In Section \ref{c09r10}, we briefly describe the scenario proposed in Rein et al.
(2010) and compare the dynamics of the simulated orbit with the best-fit solution from Correia et al. (2009). In this
section we also present the phase space portraits of the 3/2 MMR configuration of the system and the quasi-resonant
configuration. In Section \ref{phasesp}, we present dynamical maps for the region between 2/1 and 3/2 MMRs. In Section
\ref{scenario}, we describe the two stages of formation of HD\,45364 proposed by our scenario. The process of capture
in the 3/2 MMR is described in Section \ref{sectionacr}. The conclusions are presented in the last section.


\section{Scenario 1 (Rein et al. 2010).}\label{c09r10}

\begin{table*}[htbp]
\begin{center}
\begin{tabular}{|l|c|c|c|c|c|}
\hline
   &   &   &   &   & \\
 Object & parameter & C09 & R10 & $K3$ & $K100$ \\ [1.0ex]
\hline
   &   &   &   &   & \\
 HD\,45364 & Mass (M$_\odot$) & 0.82 & 0.82 & 0.82 & 0.82 \\ [2.5ex]
 HD\,45364\,\textbf{b}       & $m$ $\sin{i}$ (M$_{\rm J}$) & 0.1872 & 0.1872          & 0.1872 & 0.1872 \\
          & $a$ (AU)         & 0.6813                      & 0.6804 & 0.683 & 0.683 \\
          & $e$              & 0.17                        & 0.036  & 0.18   & 0.03 \\
          & $\lambda$ (deg)  & 105.8                       & 352.5  & 141.9  & 330.5 \\
          & $\varpi$ (deg)   & 162.6                       & 87.9   & 40.1   & 239.49\\ [2.0ex]
 HD\,45364\,\textbf{c}      & $m$ $\sin{i}$ (M$_{\rm J}$) & 0.6579 & 0.6579  & 0.6579 & 0.6579 \\
          & $a$ (AU)        & 0.8972 & 0.8994 & 0.897 & 0.905 \\
          & $e$             & 0.097 & 0.017 & 0.068 & 0.01 \\
          & $\lambda$ (deg) & 269.5 & 153.9 & 228.2 & 62.8 \\
          & $\varpi$ (deg)  & 7.4 & 292.2 & 219.9  & 58.2 \\ [2.0ex]
       & $\chi^2$ & 2.79 & 3.51 (2.76) & 3.9  & 4.2\\ [0.5ex]
\hline
\end{tabular}
\end{center}
\caption{Orbital parameters of HD\,45364 system from the best-fit of Correia et al. (2009): \textit{C09}; the F5
simulation of Rein et al. (2010): \textit{R10}; and our simulations: $K3$ and $K100$. The last row shows the corresponding $\chi^2$-values; the number in brackets for R10 corresponds to the $\chi^2$-value obtained in the case when the planet masses were not fixed.} \label{tablex1}
\end{table*}

Gravitational interactions between the protoplanetary disk and fully formed planets drive the bodies towards the
central star (Lin \& Papaloizou 1979, Goldreich \& Tremaine 1979, 1980). In the case where planet migration is
convergent (i.e. the mutual distance is decreasing), two planets can be captured into a mean-motion commensurability
and continue to migrate in the resonance lock (e.g. Lee \& Peale 2002, Ferraz-Mello et al. 2003, Beaug\'e et al. 2003,
Beaug\'e et al. 2006). This mechanism is generally accepted to explain the existence of several extrasolar systems
inside mean-motion resonances. The most populated one is the 2/1 MMR, which is a very strong first-order resonance with an extensive domain of stable motion. Due to this feature, it acts as a natural dynamical barrier which prevents most planets, formed far apart, to migrate down to other commensurabilities, such as the 3/2 MMR.

In order to understand the existence of the HD\,45364 system, Rein et al. (2010) proposed a scenario based on Type III planet migration (Masset \& Papaloizou 2003). This mode of inward migration is characterized by the very fast decay which would allow the planets to overshoot the 2/1 MMR and to be captured in the 3/2 MMR. N-body and hydrodynamic simulations showed that this is indeed possible, although it seems to require a surface density for the protoplanetary disk at least 5 times higher than the Minimum Mass Solar Nebula (MMSN, Hayashi 1981). The authors concluded that this could be the first direct evidence for Type III planet migration.

\begin{figure}
\begin{center}
\epsfig{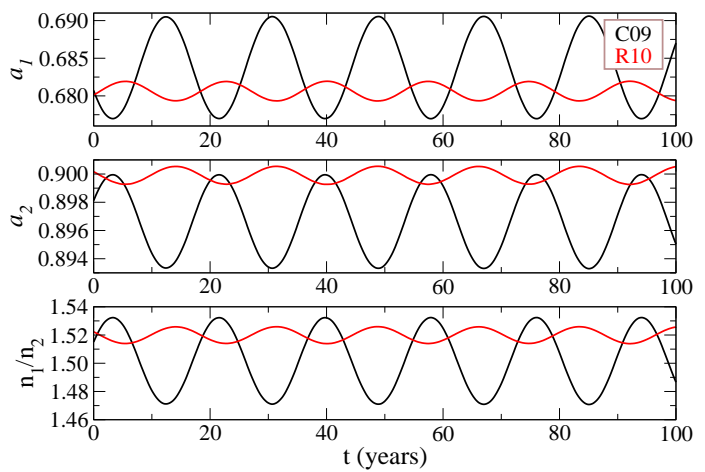} \caption{Time evolution of the semimajor axes of the inner planet (top
panel), of the outer planet (middle panel) and the mean-motion ratio (bottom panel). The black curves correspond to the best-fit C09 of Correia et al. (2009), while the red curves correspond to the solution R10 of Rein et al. (2010) }
\label{fig1}
\end{center}
\end{figure}

\begin{figure}
\begin{center}
\epsfig{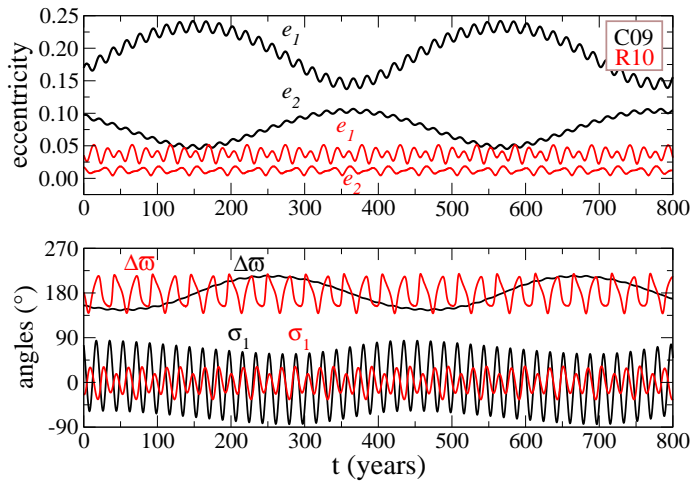} \caption{Same as in Figure \ref{fig1}, except for the planet
eccentricities (top panel) and the characteristic angles $\sigma_1$ and $\Delta \varpi$ (bottom panel).} \label{fig2}
\end{center}
\end{figure}

However, comparing the dynamics of the simulated in Rein et al. (2010) solution (hereafter, R10) with that given by
the best-fit solution in Correia et al.(2009) (hereafter, C09), we note substantial differences between them. Figures
\ref{fig1} and \ref{fig2} show the time evolution of both orbits, averaged over high-frequency terms of order of the
orbital periods. The initial orbital parameters used in the integrations are listed in columns 3 and 4 of Table
\ref{tablex1}.

Both figures show that the orbital elements of the simulated R10 system (red curves) oscillate with very small
amplitudes, in contrast with the fitted C09 system (black curves). The mean-motion ratio $n_1/n_2$ of R10 oscillates
around $1.52$, while, for C09, this occurs around $1.50$. The periods of the resonant oscillations estimated from the
semimajor axes are $T_{res} \sim  19$ years for C09, and $T_{res} \sim 17$ years for R10. A higher discrepancy is noted in the secular period, as estimated from the eccentricity variations, giving $T_{sec} \sim 400$ years for C09, and only $T_{sec} \sim 50$ years for R10. The bottom panel of Figure \ref{fig2} shows the evolution of the characteristic angles of the system, the resonant angle $\sigma_1 = 2 \lambda_1 - 3 \lambda_2 + \varpi_1 $ and the secular angle $\Delta
\varpi = \varpi_1 - \varpi_2$, where $\lambda_i$ are the mean longitudes and $\varpi_i$ are the longitudes of
pericenter; hereafter, the subscript $1$ is used for the inner planet, while $2$ is reserved for the outer body. We can see that, for R10, $\sigma_1$ oscillates with a very small amplitude when compared to its behavior in C09 and the amplitude of the secular component of $\Delta \varpi$ tends to 0.

\subsection{The Topology of the Phase Space of the C09 and R10 Systems.}\label{phasespace}

A simple comparison of the evolution of the orbital elements is insufficient to understand why both solutions are so
different. For this task, we need to visualize the phase space of the 3/2 MMR plotting energy levels of the system on
the $(e_1,e_2)$--plane of initial eccentricities. The energy ${\overline {\mathcal E}_{\rm res}}$, together with two integrals of motion, the total angular momentum ${\mathcal AM}$ and the spacing parameter ${\mathcal K}$, are given by the expressions (Michtchenko et al. 2008\,a-b):
\begin{equation}
\begin{array}{ccl}
{\overline {\mathcal E}_{\rm res}} &=& -\sum_{i=1}^2{\frac{Gm_0m_i}{2\,a_i}}-\frac{1}{2\pi}\int_0^{2\pi}\,
{\mathcal R}(a_i,e_i,\sigma_i,Q)\,dQ, \\
             & & \\
{\mathcal AM} &=& m_1\,n_1\,a_1^2\,\sqrt{1-e_1^2}+m_2\,n_2\,a_2^2\,\sqrt{1-e_2^2} {\rm ,} \\
             & & \\
{\mathcal K} &=&(p+q)\,m_1\,n_1\,a_1^2+p\,m_2\,n_2\,a_2^2{\rm ,} \label{eq3}
\end{array}
\end{equation}
where the orbital elements, including the semi-major axes $a_i$ and eccentricities $e_i$ ($i=1,2$), are canonical astrocentric variables (Ferraz-Mello et al. 2006). The averaging of the disturbing function ${\mathcal R}$ is done with respect to the synodic angle $Q=\lambda_2-\lambda_1$, where $\lambda_i$ are the mean longitudes of the planets. We fix the initial values of the resonant angles $\sigma_1$ and $\sigma_2$ at 0 or $180^0$ and calculate the values of the semimajor axes of the planets (required to calculate energy levels) using the expressions for the constants of motion ${\mathcal AM}$ and ${\mathcal K}$ (Equations \ref{eq3}).

\begin{figure}
\begin{center}
\epsfig{figure=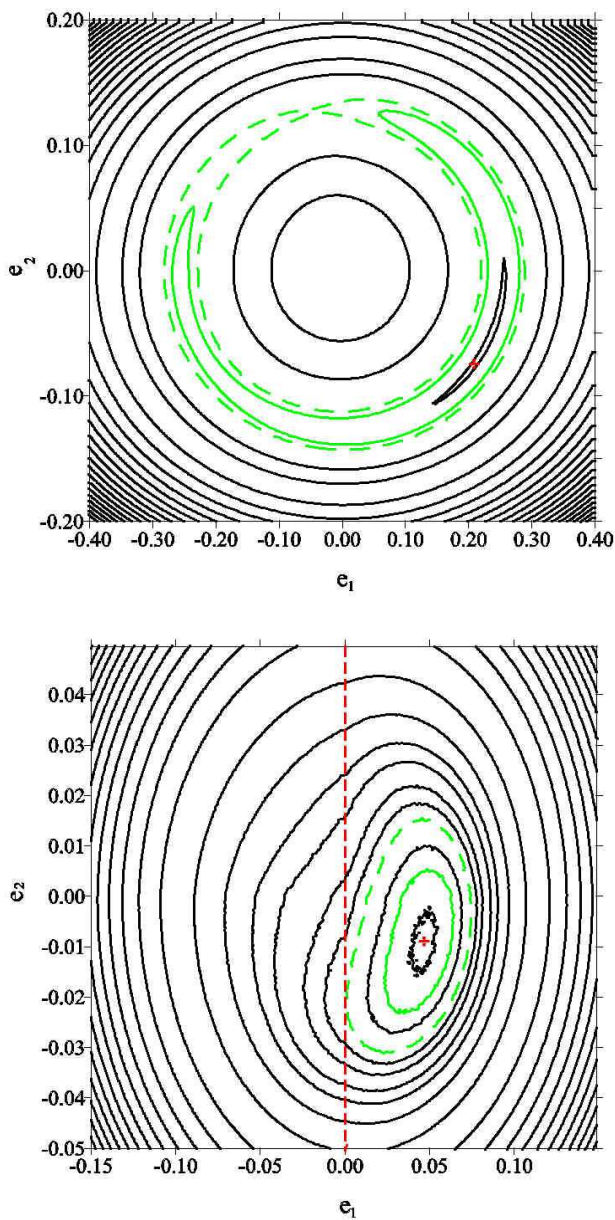,width=0.9\columnwidth,clip=} \caption{Energy levels of the 3/2 resonant Hamiltonian
(\ref{eq3}) on the ($e_1$, $e_2$)--plane, for the C09 solution (top) and for the R10 solution (bottom).
The dashed green lines show the true separatrix (top) and the kinematic transition from oscillations to circulations
(bottom). The ACR of each solution is presented by a red symbol, while the corresponding energy level by a continuous
green curve.} \label{fig16}
\end{center}
\end{figure}

Two ($e_1$, $e_2$)--planes are shown in Figure \ref{fig16}: the top graph was constructed with the parameter set of the best-fit C09 configuration, while the bottom graph corresponds to the simulated R10 system. On both graphs, the positive (negative) values on the $e_i$-axis correspond to $\sigma_i$ fixed at $0$ ($180^\circ$). The stable stationary solutions corresponding to the maximal energy of the system (frequently referred to as \emph{ACR-solutions}) are shown by  red dots in the both planes.

The comparison of dynamical structures of two planes allows us to understand the different behavior of the simulated system R10 (Rein et al. 2010) and the best-fit solution C09 (Correia et al. 2009). Indeed, although the resonant angles are oscillating in both cases, these oscillations are topologically different. In the case of R10 (bottom panel), the oscillations are merely kinematic since all levels, even that passing through the origin (dashed green curve), belong to a same structurally stable family. In the other words, there is no topological difference between oscillations and circulations. In this case, we say that the system is in a quasi-resonant regime of motion.

In contrast, for the C09 solution (top panel), the transition from oscillation of the resonant angle to circulation occurs through true bifurcations of the solutions, along the separatrix which contains a saddle-like point (dashed green curve). In the other words, both regimes of motions, oscillations and circulations, are topologically distinct. In this case, we say that the resonant angle librates and the system is in a true resonance state.

\subsection{Radial Velocity Curves.}\label{RVC}

\begin{figure}
\begin{center}
\epsfig{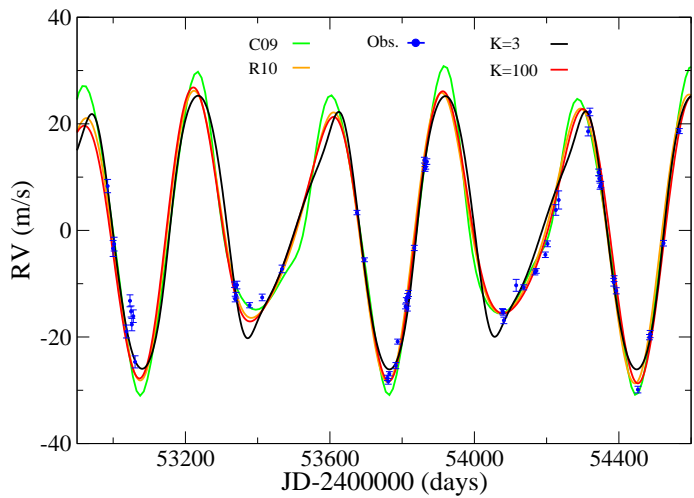} \caption{Comparison of different orbital solutions: radial velocity
measurements (blue dots), the best-fit solution C09 (green line), the R10 simulation (yellow line) and our two solutions with $K3$ (black line) and $K100$ (red line). See text for details.} \label{fig-vr}
\end{center}
\end{figure}

It is worth noting that, notwithstanding the distinct dynamical behaviour of two solutions C09 and R10, both lead to practically the same radial velocity curve. This is illustrated in Figure \ref{fig-vr}, where the observational data are shown by dots and the solutions C09 and R10 are shown by green and yellow lines, respectively. As has been noted in Rein et al. (2010), it is very difficult to see any differences in the quality of both fits; moreover, the similar  $\chi^2$-values (see Table \ref{tablex1}) suggest that two solutions are statistically indistinguishable.

\section{Dynamical Maps of the Region Between the 2/1 and 3/2 Resonances.}\label{phasesp}

\begin{figure}
\begin{center}
\epsfig{figure=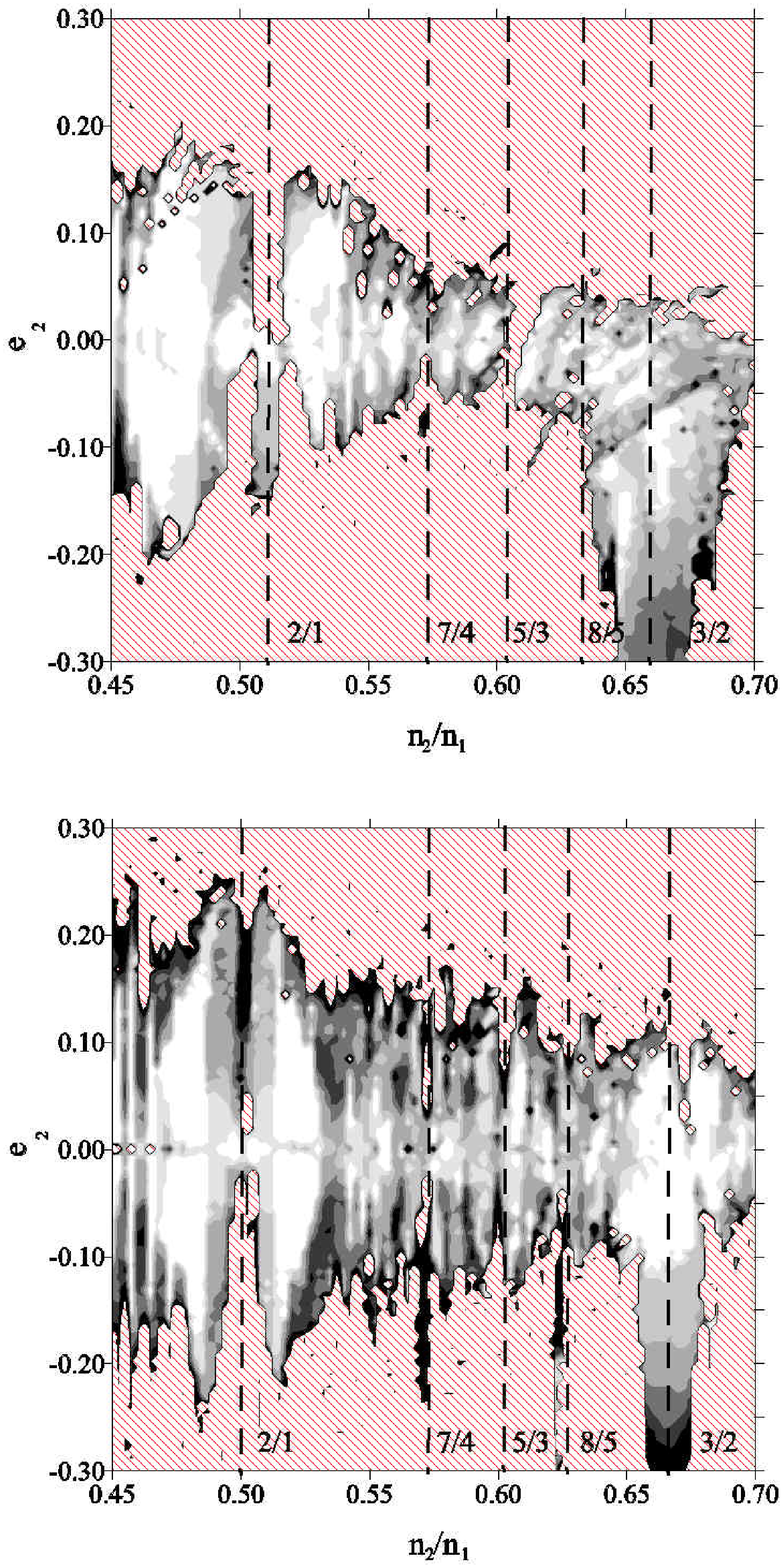,width=0.9\columnwidth,clip=} \caption{Dynamical maps of the domain between the 2/1 and 3/2
MMRs, for two fictitious systems composed of a solar mass star and two planets with masses $m_1=m_2= 1.0$\,M$_{\rm J}$
(top panel) and $m_1=m_2= 0.1$\,M$_{\rm J}$ (bottom panel). The initial conditions are $e_1=0$ and $\sigma_1=0$;
$\Delta\varpi$ is fixed at 0 (positive values on the $e_2$-axis) or $180^\circ$  (negative values on the $e_2$-axis).
The location of several MMRs is shown by vertical dashed lines. The light gray tones correspond to regions of regular
motion, while the hatched red regions correspond to strongly chaotic motion. } \label{fig3}
\end{center}
\end{figure}

To investigate the possible migration routes of the HD\,45364 system towards the 3/2 MMR, we analyzed the dynamics in
the region between the 2/1 and 3/2 MMRs. This was done using dynamical maps drawn in the $(n_2/n_1,e_2)$ representative plane. Here we only briefly describe their construction;  details can be found in Michtchenko et al. (2006a). The
$(n_2/n_1,e_2)$--plane was covered with a rectangular grid of initial conditions, with spacings $\Delta(n_2/n_1) =
0.002$ and $\Delta e_2 = 0.002$. The semimajor axis and the eccentricity of the inner planet were fixed at $a_1=1$ AU
and $e_1=0.001$, respectively. The initial values of the mean longitudes were fixed at $\lambda_1=\lambda_2=0$, while
the secular angle was fixed at $\Delta \varpi = 0$ (positive values on the $e_2$-axis in Figure \ref{fig3}) or
$180^\circ$ (negative values on the $e_2$-axis in Figure \ref{fig3}). The maps were constructed for two values for the
planetary masses; the top graph shows results for $m_1=m_2=1.0$\,M$_{\rm J}$, while in the bottom graph we used
$m_1=m_2=0.1$\,M$_{\rm J}$. These values were chosen to understand qualitatively the dependence of the dynamical
features on the individual planetary masses.

Each point of the grid was numerically integrated over $1.3 \times 10^5$ years, and the output was Fourier analyzed, in order to obtain the spectral number $N$. This quantity is defined as the number of peaks in the power spectrum of the
eccentricity of the inner planet and is used to qualify the chaoticity of planetary motion (see Michtchenko et al.
2002): small values of $N$ (light shades) correspond to regular motion, while the large values (darker tones) indicate
the onset of chaos.

The map obtained for $m_1=m_2=1.0$\,M$_{\rm J}$ is shown on the top panel in Figure \ref{fig3}. The regions of regular
motion are presented in light gray tones and the hatched red regions correspond to strongly chaotic motion leading to collisions between the planets. The domain of the 2/1 MMR is located around $n_2/n_1=0.5$; it is bounded by the chaotic layers, which correspond to the separatrix of the resonance. The robust structure of the resonance indicates that the smooth passage of the system during converging Type II migration must be interrupted, either by capture inside the resonance domain (at small eccentricities) or by ejection of the planets (at high eccentricities). There are also other mean-motion resonances between 2/1 and 3/2 MMR: they are 7/4, 5/3 and 8/5 commensurabilities located at $n_2/n_1 \sim 0.57$, $0.6$ and $0.625$, respectively. Of these, the second-order 5/3 MMR is also sufficiently strong to capture or disrupt the system during slow migration.

\begin{figure}
\begin{center}
\epsfig{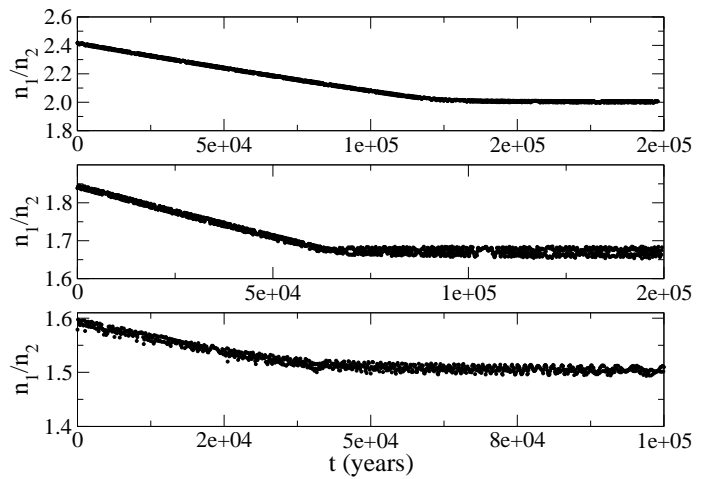} \caption{The resonance capture of a Jupiter-mass planet evolving
under the Type II migration. Starting at $n_1/n_2 > 2$, the planet is ultimately trapped in the 2/1 MMR (top panel).
Starting at $1.67 < n_1/n_2 < 2$, the planet is trapped in the 5/3 MMR (middle panel) and, starting at $n_1/n_2 <
1.67$, in the 3/2 MMR (bottom panel).} \label{fig4}
\end{center}
\end{figure}

We tested the possibility of trapping inside each of these resonances, by performing three N-body simulations. Results are shown in Figure \ref{fig4}. In all cases we considered a system of two equal mass planets ($m_1=m_2=1.0$\,M$_{\rm J}$) orbiting a Solar-mass star. The inner planet was placed at $1$ AU, while the position of the outer planet was defined by three different values of the mean-motion ratio. The initial values of the eccentricities and the angular variables of both planets were fixed at zero. The migration was simulated using a fictitious non-conservative force leading to a orbital decay of the outer planet with a characteristic timescale of $\tau_{a_2} = 10^5 $ years (Beaug\'e et al. 2006).

The first simulation was done with the outer planet starting at $n_1/n_2=2.5$, beyond the 2/1 MMR. The time evolution of the mean-motion ratio of this system (top panel in Figure \ref{fig4}) shows a capture in the 2/1 MMR at $\sim 1.3\times 10^5$ years. The system starting between 2/1 and 5/3 MMR, at $n_1/n_2=1.8$, was captured in the 5/3 MMR at $\sim 3\times 10^4$ years (middle panel). The capture in the 3/2 MMR was only possible for the systems starting between 5/3 and 3/2 MMRs, for instance, at $n_1/n_2=1.61$ (bottom panel).

It is worth noting that the captures shown in Figure \ref{fig4} are fairly robust when considering a slow
(Type II) migration. Therefore, to explain the existence of the HD\,45364 system in the context of the Type II
migration, the two planets should be formed very close each to other, in a small region located practically on the top
of the 3/2 MMR. This hypothesis, however, seems very unlikely.

\section{Scenario 2 for the Origin of the 3/2 Resonance for the HD\,45364 System.}\label{scenario}

We introduce an alternative scenario which describes the trapping in the 3/2 MMR of the embryo-size planets which are still in the initial stage of the planetary formation. It is worth noting that this possibility has been already mentioned (but not studied) in Rein et al. (2010).

Our main idea is well illustrated on the bottom panel in Figure \ref{fig3}, where we present a dynamical map analogous
to that on the top panel, except the individual planetary masses were chosen ten times smaller. Since the mutual
perturbations are weaker in this case, the size of the resonant domains and their chaotic layers decreases (Michtchenko et al. 2008 a-b). This is clearly seen in the case of the 5/3 and 2/1 MMRs  located at $n_2/n_1=0.6$ and $0.5$, respectively. As a consequence, two small planets undergone to the Type I migration process, are able to by-pass the 2/1 and 5/3 MMRs toward the 3/2 MMR.

According to current theories, tidal interactions between planets and a gas disk originate a migration process which is categorized as {\it Type I}, for planetary embryos and small core-planets ( $\leq 10.0$\,M$_\oplus$), and as {\it Type
II}, for massive giant planets. The precise delimitation between two types is defined by many factors, ones of these are the viscosity of the disk  and its scale height (Armitage 2010, Crida \& Morbidelli 2007). The migration types are qualitatively different: the Type I timescale is inversely proportional to planetary masses, while the Type II timescale is mainly determined by the physical properties of the disk. The most rapid migration is predicted to occur for the masses between $0.03$\,M$_{\rm J}$ and $1.0$\,M$_{\rm J}$ (Armitage 2010, Capt. 7).

The giant planet formation scenario provides a formation timescale of order of $8\times 10^6$ years (Armitage 2010). The mass growth is separated in three stages (for more details see Pollack et al. 1996, Rice \& Armitage 2003): core formation, hydrostatic growth and runaway growth. The formation timescale of a giant planet is almost entirely determined by the second stage, which ends when the mass of the planet envelope becomes equal to the mass of the planet core. However, the duration of the hydrostatic growth is reveled to be much longer than the typical time scales of the planet  migration. To overcome this problem, Alibert et al. (2005) suggested including the migration effects in the formation model. This scenario decreases the formation timescale to few $ 10^6$ years. It also considers only two stages of mass growth: The first stage is of core formation ceased when the planets reach the critical mass (mass of the envelope equal to mass of the core).  The second stage is of runaway growth starting when the planet masses exceed the critical mass value. The duration of the first stage of core formation is around $\sim 1 \times 10^6$ years (Alibert et al. 2005), and of the second stage of runaway mass growth is around $\sim 1 \times 10^5$ years (Armitage 2010); in such a way, the process of planet formation is completed after $\sim 1.1 \times 10^6$ years.

In this paper, we adopt the Alibert et al. (2005) scenario to explain the existence of the HD\,45364 system. We
suggest that the approach of the planet pair to the 3/2 MMR occurred during the first phase of the planet growth, when
the planet migration toward the 3/2 MMR was favored by the small masses and the fast Type I migration. Once the planets had been trapped in this resonance, they continued to grow in the stage 1, up to their masses reach $\sim$ 10\,M$_\oplus$ (a typical critical mass value), when they evolved to the stage 2 (runaway growth) and a slower Type II migration. In our experiments, this process is separated schematically into two phases which are described in the following.

\subsection{Stage I: Core Formation, Type I Migration and Capture into the 3/2 MMR.} \label{stage1}

In the first stage of the evolution, the system is composed initially of the two planetary embryos of equal masses
($0.1$\,M$_\oplus$) orbiting a central star of the mass $0.82$\,M$_\odot$ (HD\,45364). The planetary embryos
are embedded in the protoplanetary disk (we consider a vertically isothermal and laminar disk); the interactions of the planets with the disk produce their Type I convergent migration toward the central star.  This stage lasts over $ 1 \times  10^6$ years and the planets grow up to $ 10.0$\,M$_\oplus$.

The modeling of the evolution of the system in this stage is in the following. The mass growth is approximated by an exponential law $m(t)= m_\circ\,exp^{t/\tau}$, with the equal initial masses $m_\circ=0.1$\,M$_\oplus$ and the timescale $\tau=2 \times 10^5$ years, for both planets. This e-folding time provides the planet growth up to $\sim 10$\,M$_\oplus$ after $1.0 \times 10^6$ years. Although there is no widely accepted expression for the mass growth in this stage of planet formation, the exponential law seems to be a good approximation (Alibert et al. 2005, see Figures 5 and 8 in that paper); on the other hand, other authors consider different expressions, for instance, $\sin ^2 t$ (Lega et al. 2013). The chosen parameters correspond to sufficiently small planets to experiment initially the Type I migration, but also to sufficiently large planets to yield migration timescales of order of some $10^5$ years (see Armitage 2010, Capt.7). It is worth noting that, in our case, the Type I migration is still much slower than that of the Type III considered in Rein et al. (2010).

To simulate the migration during the first stage, we apply the semi-analytical models of Tanaka et al. (2002) and Tanaka \& Ward (2004), which provide the decay and damping rates of a planet, with mass $m$, semimajor axis $a$
and eccentricity $e$, orbiting a star of mass $m_*$ and embedded in a laminar disk:
\begin{equation}
\begin{array}{ccl}
\dot{a} &\simeq& - \frac{2.7 + 1.1 q}{h^2} \frac{ m }{ m_*^{1.5}} \sqrt{{\mathcal G}} \Sigma_0 a^{(1.5-q)},   \\
 &  &  \\
\dot{e} &\simeq&  \frac{e}{h^2} \frac{0.78}{(2.7+1.1 q)} \frac{\dot{a}}{a},
\label{form1}
\end{array}
\end{equation}
where ${\mathcal G}$ is the gravitational constant, and $h$, $q$ and $\Sigma_0$ are the scale height, the sharp of the surface density profile and the surface density of the disk at 1 AU, respectively. It is worth mentioning that  the model is valid only for small eccentricities ($e \lesssim 0.05$) and small planetary masses ($ \lesssim 13$\,M$_\oplus$). In the case of a smooth density profile, explicit expressions for the forces acting on the planetary masses can be found in Ogihara \& Ida (2009) and Ogihara et al. (2010), while their applications can be found in Giuppone et al. (2012). In our simulation, we assume the scale height $h=H/r = 0.05$ and the constant surface density profile $\Sigma(r) = \Sigma_0 (r/1AU)^{-q}$ with $q=0$ and $\Sigma_0=150\,g/cm^2$, which corresponds to the Minimum Mass Solar Nebula at 5 AU (MMSN, Hayashi 1981). It should be noted that Rein et al. (2010) assume the more massive protoplanetary disk, with the surface density $5$ times higher than the MMSN. From Equations (\ref{form1}), we estimate that, for the chosen disk parameters, the $e$-damping is $\sim$ 100 times faster that the $a$-decay.

\begin{figure}
\begin{center}
\epsfig{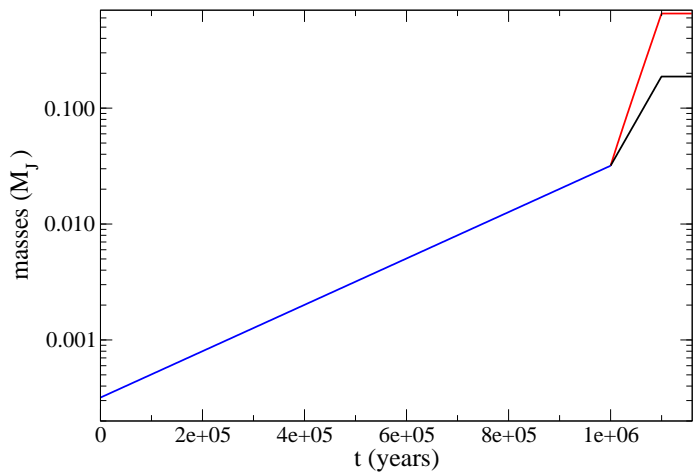} \caption{Two stages of the mass growth of the HD\,45364 planets (in logarithmic scale). During the first stage (blue line) that lasts $10^6$ yr, the equal mass planets grow with the same rate, from $0.1$\,M$_\oplus$ to $\sim 10.0$\,M$_\oplus$. During the second stage, the runaway growth occurs with different rates, in such a way that the inner planet (black line) reaches $0.187$\,M$_{\rm J}$ and the outer planet (red line) $0.658$\,M$_{\rm J}$ after $\sim 1.5 \times 10^5$ years.} \label{fig10}
\end{center}
\end{figure}

\begin{figure}
\begin{center}
\epsfig{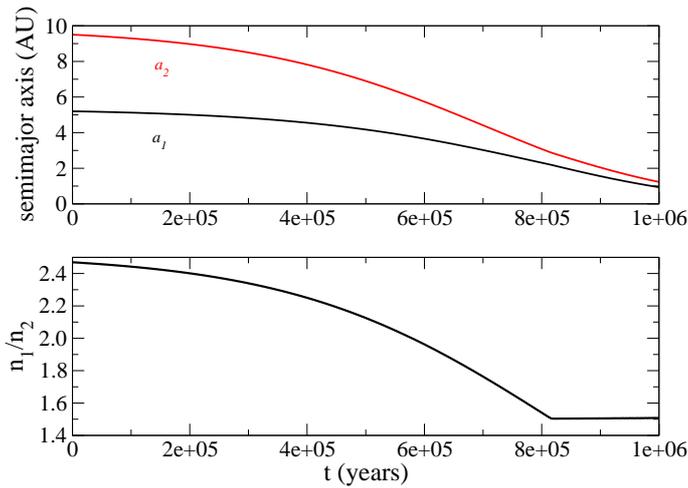} \caption{Orbital decay of the HD\,45364 planets during the first
stage of the mass growth. The top panel shows the evolution of both semimajor axes, while the bottom panel shows the
mean-motion ratio as a function of time. The system approaches the 3/2 MMR at $\sim 8 \times 10^5$ years.}
\label{fig6}
\end{center}
\end{figure}

\begin{figure}
\begin{center}
\epsfig{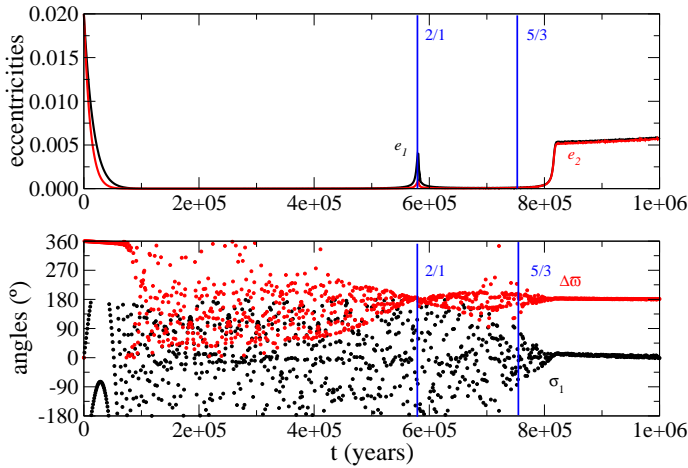} \caption{Evolution of the eccentricities (top panel) and the
angles $\sigma_1$ and $\Delta \varpi$ (bottom panel) during the first stage of the mass growth. The location of the 2/1 and 5/3 MMRs is shown by vertical lines.} \label{fig7}
\end{center}
\end{figure}

The inner planet was placed initially on the orbit similar to that of Jupiter, at $a_1=5.2$ AU, while the outer planet was located on the Saturn orbit at  $a_2=9.5$ AU, yielding the initial configuration beyond the 2/1 MMR, at
$n_1/n_2 \sim 2.46$ (or $n_2/n_1 \sim 0.4$). The initial eccentricities of the planets were $e_1=e_2=0.02$, and the
values of their angular elements were chosen equal to zero. The described set of the disk parameters and the initial
configurations will be referred hereafter to as \emph{standard configuration}; it is summarized in Table
\ref{tablex2} (first row). The results of the planet growth simulation based on the standard configuration are shown in Figures \ref{fig10}--\ref{fig7}.

The blue line in Figure \ref{fig10} shows the mass evolution (identical for both planets) during the first phase of
formation. Figure \ref{fig6} shows the decay of both semimajor axes (top graph) and of the mean-motion ratio (bottom
graph). The smooth evolution of the system, even during the passages through the 2/1 and 5/3 MMRs, is noticeable.
The system approaches the 3/2 MMR at $\sim 8 \times 10^5$ years.

The time evolution of the planet eccentricities is shown in Figure \ref{fig7}\,top. Due to the strong damping
provoked by interactions with the disk, both eccentricities drop rapidly to 0. The passage through the 2/1 MMR, at $\sim 5.7 \times 10^5$ years, provokes some visible excitations of both eccentricities, which are damped again as soon as the system leaves the 2/1 MMR. When the planets cross the strong 2/1 MMR, their masses are still small, $\sim 1.2$\,M$_\oplus$, the eccentricity excitation is weak and the resonant capture is not observed.

Approaching the very strong 3/2 MMR after $\sim 8 \times 10^5$ years of evolution, the both planets have the masses $\sim 6$\,M$_\oplus$. The excitation of the planet eccentricities (Figure \ref{fig7}\,top) and the characteristic behaviour of the resonant angle $\sigma_1$ and the secular angle $\Delta\varpi$ (Figure \ref{fig7}\,bottom), confirm that the system evolves close to the 3/2 MMR.

The first stage ceases after $ 10^6$ years, with two planets locked near the 3/2 MMR. The masses of the inner and
outer planets have grown up to  to $10.0$\,M$_\oplus$ ($\sim 0.03$\,M$_{\rm J}$). This value corresponds to
the critical value, for which the envelope mass of the planets is nearly equal to their core mass (Alibert et al. 2005, Armitage 2010) and runaway growth begins (Pollack et al. 1996, Rice \& Armitage 2003, Ikoma et al. 2000).
\begin{table*}[htbp]
\begin{center}
\begin{tabular}{| p{1.5cm} | p{0.6cm} | p{0.6cm} | p{1cm} | p{0.6cm} | p{0.6cm} | p{1.1cm} | p{1.1cm} | p{0.6cm} | p{1.5cm} |}
\hline
 & \multicolumn{3}{|c|}{ } & \multicolumn{5}{|c|}{  } & \\
 Simulation & \multicolumn{3}{|c|}{Disk parameters} & \multicolumn{5}{|c|}{Initial conditions of planets}  & Results \\ [1.0ex]
\hline
 & & & & & & & & & \\
  & \multicolumn{1}{|c|}{q} & \multicolumn{1}{|c|}{h} & \multicolumn{1}{|c|}{$\Sigma_0$} & \multicolumn{1}{|c|}{$a_1$} & \multicolumn{1}{|c|}{$a_2$} & \multicolumn{1}{|c|}{$e_1=e_2$}  &\multicolumn{1}{|c|}{$\lambda_1 = \lambda_2$}& \multicolumn{1}{|c|}{$\Delta \varpi$}& \multicolumn{1}{|c|}{$n_1/n_2 $} \\[0.4ex]
    & \multicolumn{1}{|c|}{ } & \multicolumn{1}{|c|}{ } & \multicolumn{1}{|c|}{MMSN} & \multicolumn{1}{|c|}{AU} & \multicolumn{1}{|c|}{AU} & \multicolumn{1}{|c|}{ }  &\multicolumn{1}{|c|}{deg }& \multicolumn{1}{|c|}{deg }& \multicolumn{1}{|c|}{ } \\[1.0ex]
\hline
 & & & & & & & & & \\
  \multicolumn{1}{|c|}{Standard}& \multicolumn{1}{|c|}{0} & \multicolumn{1}{|c|}{0.05} & \multicolumn{1}{|c|}{1} & \multicolumn{1}{|c|}{5.2} & \multicolumn{1}{|c|}{9.5} & \multicolumn{1}{|c|}{0.02}  &\multicolumn{1}{|c|}{0}& \multicolumn{1}{|c|}{0}& \multicolumn{1}{|c|}{$ 3/2$} \\[2.0ex]
    \multicolumn{1}{|c|}{1}& \multicolumn{1}{|c|}{0} & \multicolumn{1}{|c|}{0.05} & \multicolumn{1}{|c|}{1} & \multicolumn{1}{|c|}{5.2} & \multicolumn{1}{|c|}{9.0} & \multicolumn{1}{|c|}{0.02}  &\multicolumn{1}{|c|}{0}& \multicolumn{1}{|c|}{0}& \multicolumn{1}{|c|}{$ 3/2$} \\
  \multicolumn{1}{|c|}{2} & \multicolumn{1}{|c|}{0} & \multicolumn{1}{|c|}{0.05} & \multicolumn{1}{|c|}{1} & \multicolumn{1}{|c|}{5.2} & \multicolumn{1}{|c|}{9.5} & \multicolumn{1}{|c|}{0.04}  &\multicolumn{1}{|c|}{0}& \multicolumn{1}{|c|}{0}& \multicolumn{1}{|c|}{$ 3/2$} \\
    \multicolumn{1}{|c|}{3}& \multicolumn{1}{|c|}{0} & \multicolumn{1}{|c|}{0.05} & \multicolumn{1}{|c|}{1} & \multicolumn{1}{|c|}{5.2} & \multicolumn{1}{|c|}{9.5} & \multicolumn{1}{|c|}{0.02}  &\multicolumn{1}{|c|}{$\lambda_1 \neq \lambda_2 $}& \multicolumn{1}{|c|}{0}& \multicolumn{1}{|c|}{$ 3/2$} \\
        \multicolumn{1}{|c|}{4}& \multicolumn{1}{|c|}{0} & \multicolumn{1}{|c|}{0.05} & \multicolumn{1}{|c|}{1} & \multicolumn{1}{|c|}{5.2} & \multicolumn{1}{|c|}{9.5} & \multicolumn{1}{|c|}{0.04}  &\multicolumn{1}{|c|}{0}& \multicolumn{1}{|c|}{60}& \multicolumn{1}{|c|}{$ 3/2$} \\ [1.0ex]
     \multicolumn{1}{|c|}{5}& \multicolumn{1}{|c|}{0.25} & \multicolumn{1}{|c|}{0.05} & \multicolumn{1}{|c|}{1} & \multicolumn{1}{|c|}{5.2} & \multicolumn{1}{|c|}{9.5} & \multicolumn{1}{|c|}{0.04}  &\multicolumn{1}{|c|}{0}& \multicolumn{1}{|c|}{0}& \multicolumn{1}{|c|}{$ 3/2$} \\
    \multicolumn{1}{|c|}{6}& \multicolumn{1}{|c|}{0.5} & \multicolumn{1}{|c|}{0.05} & \multicolumn{1}{|c|}{1} & \multicolumn{1}{|c|}{5.2} & \multicolumn{1}{|c|}{9.5} & \multicolumn{1}{|c|}{0.04}  &\multicolumn{1}{|c|}{0}& \multicolumn{1}{|c|}{0}& \multicolumn{1}{|c|}{ $2/1$ or $ 3/2$} \\
   \multicolumn{1}{|c|}{7}& \multicolumn{1}{|c|}{1.5} & \multicolumn{1}{|c|}{0.05} & \multicolumn{1}{|c|}{1} & \multicolumn{1}{|c|}{5.2} & \multicolumn{1}{|c|}{9.5} & \multicolumn{1}{|c|}{0.04}  &\multicolumn{1}{|c|}{0}& \multicolumn{1}{|c|}{0}& \multicolumn{1}{|c|}{ $>2.9$} \\[1.0ex]
        \multicolumn{1}{|c|}{8}& \multicolumn{1}{|c|}{0} & \multicolumn{1}{|c|}{0.07} & \multicolumn{1}{|c|}{1} & \multicolumn{1}{|c|}{5.2} & \multicolumn{1}{|c|}{9.5} & \multicolumn{1}{|c|}{0.04}  &\multicolumn{1}{|c|}{0}& \multicolumn{1}{|c|}{0}& \multicolumn{1}{|c|}{$ 3/2$} \\
        \multicolumn{1}{|c|}{9}& \multicolumn{1}{|c|}{0} & \multicolumn{1}{|c|}{0.05} & \multicolumn{1}{|c|}{2} & \multicolumn{1}{|c|}{5.2} & \multicolumn{1}{|c|}{9.5} & \multicolumn{1}{|c|}{0.04}  &\multicolumn{1}{|c|}{0}& \multicolumn{1}{|c|}{0}& \multicolumn{1}{|c|}{$ 3/2$} \\[1.0ex]
\hline
\end{tabular}
\end{center}
\caption{Simulations performed with different orbital and disk parameters (see Section \ref{stage1-1} for more details).} \label{tablex2}
\end{table*}

\subsection{Probability of the Trapping in the 3/2 MMR During the Stage I.}\label{stage1-1}

We have tested different sets of the parameters of the disk and the initial configurations of the planets. Some of these sets are present in Table \ref{tablex2}, where the first row shows the parameters of the standard configuration and the other rows show one-by-one (systematic) parameter changes of this configuration. The last column shows the final configurations of the system at the end of the stage I of evolution. We have detected three possible final configurations: the trapping in the 3/2 MMR, the capture in the 2/1 MMR and the non-resonant configuration due to the diverging evolution of the planet orbits.

From Table \ref{tablex2}, we observe that the 3/2 trapping of the standard configuration of the HD\,45364 system is not sensible to the variations of the initial conditions (runs \#1--\#4). This result may be easily understood analyzing the evolution of the orbital elements in Figures \ref{fig6} and \ref{fig7}. Indeed, due to continuous decay of the planets, the different initial positions from the central star will only accelerate/desaccelerate the trapping in the 3/2 MMR. The non-zero initial eccentricities are rapidly damped to zero-values by the disk-planet interactions, thus having no effect on the capture into a mean-motion resonance. Finally, for nearly circular orbits, the angular variables $\sigma_1$  and $\Delta\varpi$ are high-frequency circulating angles, acquiring any value from the range between 0 and $360^\circ$.

The picture is changed when we consider different disk properties (runs \#5--\#9 in Table \ref{tablex2}), specially the different values of the sharp of the surface density profile $q$. Indeed, from Equations (\ref{form1}), we have that $\dot{a} \propto a^{(1.5-q)} $. For small $q$--values (in our runs, $< 0.25$),  the decay of the outer planet is much faster than that of the inner planet ($\dot{a}_2 \gg \dot{a}_1 $), that provokes a rapid convergence of the planet orbits and the non-capture passage through the 2/1 MMR.

For larger $q$--values (run \# 6, with $q=0.5$), the convergence of the orbits becomes slower ($\dot{a}_2 > \dot{a}_1 $) and the capture inside the 2/1 MMR becomes possible, reducing the probability of the trapping in the 3/2 MMR. This result is in an agreement with the investigations of the probability of capture inside the first-order resonances as a function of the rate of the orbital convergence present in Mustill \& Wyatt (2011). Finally, for $q = 1.5$ (run \# 7), $\dot{a}_2 = \dot{a}_1 $ and the two orbits become divergent, i.e. the mutual planet distance increases during the orbital decay. 

We can conclude that, for sufficiently small $q$-values ($q \leq 0.25$), i.e. in the disk with smooth profile of surface density, the trapping inside the 3/2 MMR is robust during the first stage of the planet formation/migration.

\subsection{Stage II: Runaway Mass Growth and Type II Migration.}

The second stage of the evolution begins when the envelope mass of the planet becomes equal or larger than the core
mass of the planet; as a consequence, the rate of accretion is accelerated dramatically and the runaway growth is
initiated (Pollack et al. 1996, Ikoma et al. 2000, Armitage 2010). This phase lasts an average of approximately $\sim 1 \times 10^5$ years (Alibert et al. 2005, Armitage 2010, Hasegawa \& Pudritz 2012). During this stage, the mass growth of the planets is approximated by an exponential law $m(t)= m_i \,exp^{t/\tau_i}$, where $m_i$ are the initial masses of the planets and $\tau_i$ are the e-folding times of the planet growth ($i=1,2$). The values of $m_i$ come from the stage I and are $10.0$\,M$_\oplus$, for both planets. The values of $\tau_i$ are chosen as $5.6 \times 10^4$ years and $3.5 \times 10^4$ years, for the inner and outer planets, respectively, such that the planets reach their current masses ($0.1872$M$_{\rm J}$ and $0.6579$M$_{\rm J}$) after $1 \times 10^5$ years.

Figure \ref{fig10} shows the exponential accretion of the planet masses during the second phase: the rapid growth of
the outer planet by the red line and slower growth of the inner planet  by the black line. The growth process is
stopped after $\sim 1.1 \times 10^6$ years and the final values of the planetary masses are compatible with
those of the C09 and R10 solutions.

As the planets  become more massive, the structure of the disk in their neighborhood is modified and gaps begin to
form. From this stage onward we assume a Type II of migration of both planets already locked near the 3/2 MMR resonance. To simulate this migration, we apply a Stokes-like non-conservative force (Beaug\'e et al. 2006). The effects of the
Stokes force on the semimajor axis and eccentricity evolution are a decay and a damping, respectively, given as
\begin{equation}
a(t) = a_0 \exp (-t/\tau_a),\,\,\,\,\,\, e(t) = e_0 \exp (-t/\tau_e), \label{eq:stage2}
\end{equation}
where $a_0$ and $e_0$ are the initial values of the semimajor axis and the eccentricity, respectively, and $\tau_a$ and $\tau_e$ are the e-folding times of the corresponding orbital elements (Beaug\'e \& Ferraz-Mello 1993, Gomes 1995). The configurations of the planets at the end of the stage I of evolution were used as input of the second stage.

\begin{figure}
\begin{center}
\epsfig{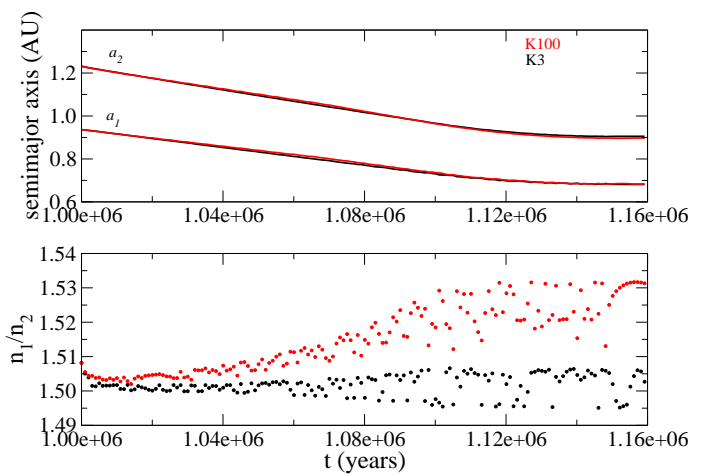} \caption{Orbital decay of the HD\,45364 planets during the second
stage of the planet formation. The top panel shows the evolution of both semimajor axes, while the bottom panel shows
the mean-motion ratio as a function of time.
The black dots correspond to the $K3$ solution and the red dots to the $K100$ solution.}
\label{fig9}
\end{center}
\end{figure}

The lost of the orbital energy during the migration of the planets toward the central star defines the choice
of the values of the e-folding times $\tau_a$. They are generally obtained through adjustment between the duration of
the migration and the final positions of the planets, which, in the case of the HD\,45364 system, are roughly
$a_1 \sim 0.68$\,AU and $a_2 \sim 0.90$\,AU (Table \ref{tablex1}). If we assume that the second stage lasted $1.5 \times 10^5$ years, we obtain $\tau_{a_1} =  6 \times 10^5$ and $\tau_{a_2} = 3 \times 10^5$ years.

\begin{figure}
\begin{center}
\epsfig{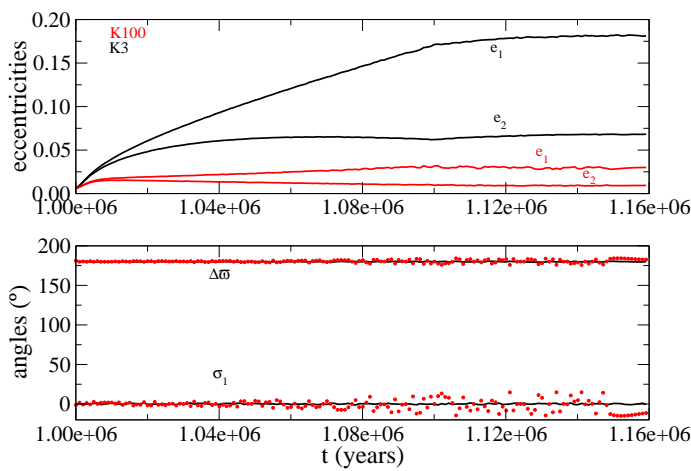} \caption{Evolution of the eccentricities (top panel) and the
angles $\sigma_1$ and $\Delta \varpi$ (bottom panel) during the second stage of the planet formation, when the planets
increase their masses from $10.0$\,M$_\oplus$ to the current values.
The black dots correspond to the $K3$ solution and the red dots to the $K100$ solution.}
\label{fig8}
\end{center}
\end{figure}

The choice of the e-folding times of the eccentricity damping is more complicated. This is because the exchange of the angular momentum between the planets and the disk is very sensitive to the properties of the disk and the
processes involved. In the simplified simulations based on the Stokes-like forces, this is usual to introduce the
factor $K=\tau_a/\tau_e$, which is a ratio of the e-folding times of the orbital decay and the eccentricity damping,
and use it as a free parameter (Lee \& Peale 2002, Beaug\'e et al. 2006, Kley et al. 2005). The typical values of the
factor $K$ lie in the range between 1 and 100 (Lee \& Peale 2002). We have tested several values of $K$ from this interval and have chosen two specific values; they are same for both planets and are 3 and 100. Hereafter, we refer to these solutions as $K3$ and $K100$. The values of the masses and the orbital configurations are given in the two last columns in Table \ref{tablex1}.

The stability of both solution was tested and confirmed using the stochasticity indicator MEGNO (Cincotta \& Sim\'o 2000). We have found that both systems achieve $<Y> =$\,2 in $\sim$ 3000 years. Both simulations are also in good agreement with the observation data. The corresponding radial velocity fits are shown in Figure \ref{fig-vr} by  black and red lines, for $K3$ and $K100$, respectively. The $\chi^2$--values shown in Table \ref{tablex1} confirm the statistical equivalence of our solutions with the solutions C09 (green line) and R10 (yellow line). It should be stressed that no effort was done to minimize the $\chi^2$--value through the fine adjustment of the parameters of the problem.

The dynamics of the $K3$ solution is similar to the best-fit C09 solution, while the $K100$ solution reproduces well the solution R10. The time evolution of the solutions $K3$ and $K100$ is shown in Figures \ref{fig9} and \ref{fig8} by black and red lines, respectively. The migration process characterized by decay of the planetary semimajor axes is shown in Figure \ref{fig9}\,top: the timespan of the decay, defined by the value of $\tau_{a_1}$ and $\tau_{a_2}$, is extended up to $\sim 1.15\times 10^6$ years. We can see that the averaged magnitudes of the semimajor axes of $K3$ at the end of migration are compatible with those of the best-fit of Correia et al. (2009) and of $K100$ with the $F5$ simulation of Rein et al. (2010) (see Table \ref{tablex1}). Although, the top graph in Figure \ref{fig9} seems to show no difference between the values of the semimajor axes of both solutions, in the enlarged scale, we can observe a small but important difference (see Table \ref{tablex1}), which implies that, for  $K=100$, the mean-motion ratio $n_1/n_2 \sim 1.52$ (close to the R10 solution); for $K=3$,  $n_1/n_2 \sim 1.5$ (close to the C09 solution).

The behavior of the eccentricities and the characteristic angles $\sigma_1$ and $\Delta \varpi$  during the second
stage of migration is shown in Figure \ref{fig8}, again by a red line for $K=100$ and a black line for $K=3$. Since the migration is sufficiently slow, the evolution of the eccentricities (top panel) follows closely the family of
stationary solutions of the conservative 3/2 resonant problem, the ACR--family (Beaug\'e et al. 2003, Michtchenko et al. 2006b).  The different rates of the mass accretion, when the outer planet grows more rapidly, are responsible for the excitation and the continuous increase of the eccentricity of the less massive inner planet (the evolutionary paths are discussed in Sect. \ref{sectionacr}). We can see that the results obtained for $K=3$ reproduce the high eccentricities of C09, while the results obtained with $K=100$ reproduce the low-eccentricity solution R10.

The evolution of the characteristic angles $\sigma_1$ and $\Delta\varpi$ (bottom panel in Figure \ref{fig8}) is
low-amplitude oscillations around $0$ and $180^\circ$, respectively, for both solutions. Despite the apparently similar behaviour of the characteristic angles in both cases, the dynamical analysis shows that the solution $K3$ is involved
deeply inside the 3/2 MMR during this stage, while the solution $K100$ is in a quasi-resonant state, similar to the
system R10 (see Sect. \ref{phasespace}).

During the last $5 \times 10^4$ years of the process, the gas disk is dissipated, while the pair of already formed
planets, trapped in the 3/2 MMR (see Figure \ref{fig9}), continues to decay slowly (since the characteristic
frequencies of the system evolution are, at least, one order higher, this process can be still considered as adiabatic). The disk is dissipated completely after $1.15\times 10^6$ years and the planets begin to evolve solely due to their mutual gravitational perturbations (last $10^4$ years in our simulation). The timespan of $1.15\times 10^6$ years is compatible with lifetime of protoplanetary disks, estimated to lie between 1-10 Myr (Armitage 2010, Haisch et al. 2001).

\section{Evolutionary Paths During Capture in the 3/2 MMR.}\label{sectionacr}

\begin{figure}
\begin{center}
\epsfig{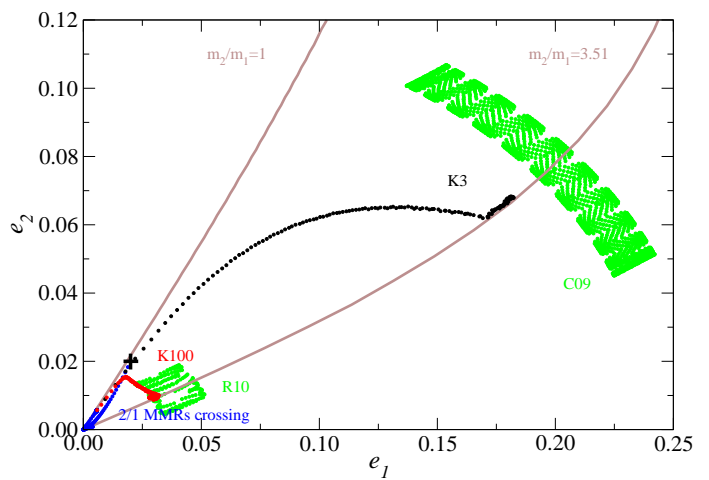} \caption{Projections of the C09 and R10 oscillations
(green dots) and  the $K3$ and $K100$ evolutionary paths, during the two stages of planet formation/migration,
on the $(e_1,e_2)$--plane. The first stage of accretion includes the passage through the 2/1 MMR and the capture
into the 3/2 MMR (blue points). The evolution during the second stage of the run-away mass growth and the Type II migration is shown by
black dots for the $K3$ solution and red dots for the $K100$ solution. The ACR-families are shown by continuous lines; one is parameterized by $m_2/m_1=1$ and other by $m_2/m_1=3.51$. The black cross symbol shows the initial configuration of the planet pair.} \label{fig14}
\end{center}
\end{figure}

Figure \ref{fig14} shows the projections of the high-eccentricity C09 solution and the low-eccentricity R10 solution
(green dots) on the $(e_1,e_2)$ -- plane. We plot two ACR--families parameterized by the ratios of the initially equal
masses $m_2/m_1=1.0$ and  of the actual planet masses $m_2/m_1=3.51$ (continue curves). The motion of the solutions C09 and R10 on this plane is an oscillation around an ACR defined by the total angular momentum ${\mathcal AM}$ and the scaling parameter ${\mathcal K}$ (see Equations (\ref{eq3})). Since the planet masses and the semimajor axes are similar, the ${\mathcal K}$--parameter is same for both solutions, and the difference in the positions of the corresponding ACR is due to distinct values of the angular momentum ${\mathcal AM}$ (or eccentricities). It must be kept in mind that the difference between two solution is qualitative: the solution C09 is resonant, while the solution R10 is quasi-resonant (Sect. \ref{phasespace}).

The same $(e_1,e_2)$--plane is also adequate to visualize the process of capture of the system inside the 3/2 MMR. We
show the projections of the two evolutionary paths, $K3$ and $K100$, during the two stages described in Sect. \ref{scenario}.

The first stage is the fast Type I migration of the initially equal mass planetary embryos ($0.1$\,M$_\oplus$) growing with a same rate up to $\sim 10.0$\,M$_\oplus$ during $ 10^6$ years; the evolution of the system during this stage is similar for both simulations. The initial planet configuration is shown by a black cross symbol, at $e_1=e_2=0.02$. Blue dots show first the rapid eccentricity damping to the circular orbits, then the passages through 2/1 MMR, with slight excitation of the planet eccentricities, and finally the approach to the 3/2 MMR. Since the planet masses are still equal during this stage, the evolution occurs along the ACR--family parameterized by $m_2/m_1=1$.

In the second stage of the slow Type II migration and the fast runaway mass accretion, the evolution of two solutions becomes distinct due to the different rates of the mass growth. Both solutions diffuse in direction of the ACR--family parameterized by the actual ratio $m_2/m_1=3.51$, but the path on the ($e_1,e_2$) done by the $K100$ solution (red dots) is much shorter than that done by the $K3$ solution (black dots). This is because the value $K=100$ implies a strong damping of the eccentricities during the second stage of the evolution and the consequent formation of the system in the quasi-resonant state, similar to that proposed by Rein et al. (2010). Thus, our scenario is able to reproduce well the R10 solution, without to invoke unusual the type III migration and high disk density.

The solution obtained with $K=3$ exhibits a resonant dynamical behaviour similar to that of the C09 configuration of the HD\,45364 system (Correia et al. 2009). Moreover, the magnitudes of the orbital elements are in agreement with the averaged values of the solution C09, also as the resonant and secular frequencies. The discrepancy with C09 we have noted is in the amplitudes of the orbital elements oscillation around the corresponding ACR solution of the maximal energy. This could be explained by simplifications introduced in our model, which, for instance, do not account for stochastic processes during the planet formation/migration. The stochastic forces can be responsible for small fluctuations of the energy and a consequent deviation of the system from the maximal energy equilibrium configuration. This would imply in the increasing oscillation amplitudes of the orbital elements of the planets observable in the best-fit solution of Correia et al. (2009).


\section{Conclusions.}\label{conclus}

In this paper we have presented a new scenario for the origin of HD\,45364, which, depending on the free parameter $K$, reproduces both the resonant and quasi-resonant configurations of the system.  The former one is associated to the best-fit solution by Correia et al. (2009) and the last one is the simulation of Rein et al. (2010). Our scenario
considers that the migration process of the planets began during the first stages of mass growth. The initial
planetary masses were chosen of the order of planetary embryos and, at the end of the formation process, they
achieved the giant planet masses. The formation of the system involved both Type I and Type II migration sets. The total time span of simultaneous  accretion and migration was assumed to have lasted $\sim 10^6$ years,
that is compatible with the lifetimes observed for protoplanetary disks.

Our scenario presents several differences with respect to that proposed by Rein et al. (2010). The first one is with respect to the type of migration assumed to have dominated the resonance trapping: instead of the non-conventional Type III migration, we consider the combined effects of the Type I and the Type II migration processes, depending on the
planetary masses. We have found that the passage across the 2/1 and 5/3 MMRs, without capture inside them, is possible during the Type I migration assuming small planetary masses instead of Jupiter-like masses. We show that this process is robust if we consider the vertically isothermal and laminar disk with the nearly constant surface density profile.

The other difference between two scenarios is with respect to the disk properties. To simulate the Type III migration, Rein et al. (2010) needed to consider  the massive protoplanetary disk, with the surface density $5$ times higher than the MMSN. On the other hand, the observational data indicate the existence of the correlation between the mass of the disk and the mass of the central star: the median disk-to-star mass ratio is $\sim 0.5\%$ (Andrews \& Williams 2005). Therefor, for the HD\,45364 star of  $0.82$\,M$_\odot$, the assumption of the disk with 5 MMSN seems to be no reasonable. Moreover, the massive disk implies in extremely short timespans of the planet formation, of order of $10^3$ years. In contrast, our scenario assumes the typical MMSN for the disk in the first stage of the planet formation/migration and the duration of the planet formation of order of $10^6$ years, compatible with the lifetimes observed for the protoplanetary disks.

We have also explored the different features of the disk during the second stage of the runaway planet formation and the slow Type II migration. This has been done introducing a free parameter $K=\tau_a / \tau_e$, defined as the ratio
between the e-folding times of the orbital decay and the eccentricity damping (Lee \& Peale 2002, Beaug\'e et al. 2006, Kley et al. 2005). We have found that, for large values of $K$, the intense eccentricity damping makes impossible the resonant capture of the planets; in such a way, the pair remains close to the 3/2 MMR, in a quasi-resonant state. It is interesting note that this feature could explain the abundance of the low-eccentricity Kepler objects close to the
strong MMRs of the first order. Our simulation done with $K=100$ produced the quasi-resonant pair of the HD\,45364 planets, similar to the F5 simulation of Rein et al. (2010).

For small values of $K$, the interactions with the protoplanetary disk allow the effective locking of the planetary motions in the 3/2 MMR, \emph{inside} the resonant domain. Our simulation done with $K=3$ produced the resonant pair of the HD\,45364 planets, similar to the best-fit simulation of Correia et al. (2009). Thus, the quasi-resonant or resonant configurations of the HD\,45364 system have shown to be a direct consequence of the properties of the gas disk assumed. The additional radial velocities measurements can confirm the real configuration  of the HD\,45364 system, allowing to assess of the physical properties of the primordial protoplanetary disk of this system.

Finally, it is worth emphasizing that the model of formation and capture of the HD45364 system in the 3/2 MMR presented here suffers from several
simplifications and limitations; for instance, the exponential law for the mass growth was assumed arbitrarily, and the complex processes produced by interactions between the gas/dust disk and the growing planets were schematized assuming simple analytical expressions for the orbital decay and eccentricity damping. On the other hand, it should be noted that i) these expressions are provided by solid theories concerning the formation and early evolution of the planets; ii) the values of physical and orbital parameters used in simulations are consistent with those observable and widely acceptable; iii) obtained final solutions are robust in the range of the typical input parameters of the model. In this way, we consider our model as a plausible mechanism to explain the possible near 3/2 resonance configuration of the HD\,45364 system.

\section*{Acknowledgments}

This work was supported by the S\~ao Paulo State Science Foundation, FAPESP, the Brazilian National Research Council,
CNPq, and the Argentinean Research Council, CONICET. This work has made use of the facilities of the Computation Center of the University of S\~ao Paulo (LCCA-USP) and of the Laboratory of Astroinformatics (IAG/USP, NAT/Unicsul), whose purchase was made possible by the Brazilian agency FAPESP (grant 2009/54006-4) and the INCT-A.  The authors are grateful to Dr. Hanno Rein for numerous suggestions/correction on this paper.



\end{document}